\documentclass[letterpaper]{article}
\usepackage{iccc}
\usepackage{todonotes}
\usepackage{comment}
\usepackage{times}
\usepackage{helvet}
\usepackage{courier}
\usepackage{url}
\usepackage{subfig}

\title{Data-driven Design: A Case for Maximalist Game Design \\
Paper type: Position paper}
\author{Gabriella A. B. Barros$^{1}$, Michael Cerny Green$^{1}$, Antonios Liapis$^{2}$ and Julian Togelius$^{1}$ \\
1: Tandon School of Engineering,
New York University,
New York, USA\\
2: Institute of Digital Games,
University of Malta,
Msida, Malta\\
gabbbarros@gmail.com, mcgreentn@gmail.com, antonios.liapis@um.edu.mt, julian@togelius.com
}
\begin{document} 
\maketitle
\begin{abstract}
\begin{quote}
Maximalism in art refers to drawing on and combining multiple different sources for art creation, embracing the resulting collisions and heterogeneity. This paper discusses the use of maximalism in game design and particularly in data games, which are games that are generated partly based on open data. Using Data Adventures, a series of generators that create adventure games from data sources such as Wikipedia and OpenStreetMap, as a lens we explore several tradeoffs and issues in maximalist game design. This includes the tension between transformation and fidelity, between decorative and functional content, and legal and ethical issues resulting from this type of generativity. This paper sketches out the design space of maximalist data-driven games, a design space that is mostly unexplored.
\end{quote}
\end{abstract}

\section{Introduction}\label{sec:introduction}

The unprecedented availability of digital data impacts most human endeavors, including game design. In particular, freely available data can be combined with procedural content generation (PCG) and computation creativity to create systems that can generate games (or game content) based on open data. We have previously identified such games as ``data games''~\cite{gustafsson2013data}.

This paper explores some of the aesthetic challenges, particularities and concerns associated with games that are created from data. We start from the idea that the use of data games is in many ways similar to notions in art such as collage, sampling, and remixing. We draw on content from many different sources, causing creative collisions between them. This lets us apply some of the same conceptual apparatus to study data games as has been applied to these types of art. We also start from a series of game generators we have created, collectively referred to as ``Data Adventures''. These generators create adventure games, such as murder mysteries, from open data from e.g. Wikipedia and OpenStreetMap. Our ongoing struggle with getting these generators to produce playable and interesting content from something as varied and occasionally unreliable as Wikipedia has illuminated both possibilities and pitfalls of this approach.


This paper is an attempt to explore the design space of maximalist data-driven games (and other data games) in order to form an initial understanding of it. It is also an attempt to systematize reflections from our own and others' attempts at creating such games. 
We address the following questions:
\begin{itemize}
\item What does it mean for games designed from/for data to be maximalist?
\item What is the tradeoff between transforming data and staying true to the source in terms of generating games?
\item What are the characteristics of game content that can be generated from data?
\item For what purposes can data-driven maximalist games be designed and how does that affect their character?
\item What new legal and ethical issues, including copyright issues and the potential for generating offensive, misinforming and biased content, are raised by this type of game design?
\end{itemize}

\section{Data-driven design and data games}

This age of data sharing (whether sharing is free or not) has certainly been advantageous to research in computational creativity. While computational creativity does not necessarily need to emulate human creativity \cite{pease2011impact}, freely available human-annotated data can be exploited as an inspiring set \cite{ritchie2007criteria} to any creative software. In natural language generation, Google N-grams have been exploited to identify analogies and similes \cite{veale2014breakingbad}, corpora of phonetic information for all words have been exploited to generate jokes \cite{ritchie2011standup}, and books of a specific author have been used to generate stories typical of the genre \cite{khalifa2017deep}. In visual generation, crowdsourced annotations of data were used to create image filters \cite{heath2016see}, while object recognition models based on deep learning of Google images was used to choose how generated 3D shapes would represent an object \cite{lehman2016objects}. Similarly, deep learning from massive musical corpora was used to create new music \cite{hawthorne2017piano}.

In the creative domain of games, on the other hand, similar approaches have been used to create different game components. Google's autocomplete function (which uses a form of N-grams) was used to discover names for enemies and abilities of a game character whose name was provided by the player \cite{cook2014rogue}. In the same game, Google image search used discovered names to select images for these enemies' sprites. In other work, \citeauthor{guzdial2016videos} (\citeyear{guzdial2016videos}) used Youtube playthroughs to find associations in the placement of level elements (e.g. platforms, enemies) which were used to generate levels for \emph{Super Mario Bros} (Nintendo 1985). Patterns in \emph{Starcraft II} maps (Blizzard Entertainment 2010) were learned through deep learning \cite{lee2016predicting}; these encodings were used to change the frequency of minerals in the map without the usual exploratory process of e.g. an evolutionary algorithm. To better coordinate the learning process of level patterns, a corpus of diverse games has been collected \cite{summerville2016corpus}. 

While using existing game data ---often annotated with human notions of quality--- has been explored in computational game creativity \cite{liapis2014computational}, most efforts perform minor adjustments to existing games. Game generators such as Angelina \cite{cook2012aesthetic}, A Rogue Dream \cite{cook2014rogue}, and Game-O-Matic \cite{treanor2012gameomatic} use data outside the game domain, enhancing their outcomes with human-provided associations (and content such as images). Even so, the core gameplay loop is simple: in Angelina, for example, the player performs the basic actions of a platformer game (e.g. jump, run); in A Rogue Dream the player moves along 4 directions and perhaps uses one more action. Gameplay in all these games is mechanics-heavy, relying on fast reactions to immediate threats rather than on high-level planning or cognitive ability. Many \emph{data games} take an existing game mechanic and generate new content for that game from open data~\cite{gustafsson2013data,friberger2012monopoly,cardona2014open}. In some cases, such as the game \emph{Bar Chart Ball}, a new game mechanic is added to an existing data visualization~\cite{togelius2013barchartball}. To play even simple data games, the player must have some understanding of the underlying data. Playing data games requires some mental effort, deduction or memory; not only dexterity.

While most data-driven game generation software focus on a simple and tight gameplay loop, there is considerable potential in using and re-using information outside of games to create more complex game systems and more involved experiences. 
We argue that data-driven game generation can allow for a new gameplay experience. Using the Data Adventures series of game generators as a concrete example, we articulate the tenets of maximalism in game design inspired by the art movement of the same name. Moreover, we discuss two possible dimensions of maximalist game design, and how it can start from the raw data on one end or from the gameplay experience on the other. Finally, we envision the potential uses and issues of maximalist game design.


\section{Maximalism in data-driven design}

We are inspired by the notion of maximalism in the arts, rather than in the game design sphere. In music, for example, maximalism ``embraces heterogeneity and allows for complex systems of juxtapositions and collisions, in which all outside influences are viewed as potential raw material'' \cite{jaffe1995maximalist}. We similarly embrace the use of heterogeneous data sources as notes (i.e. the individual components) and melody (i.e. the overarching game or narrative structure) to produce a game as an orchestration of dissimilar instruments \cite{liapis2015orchestration}. In that sense, maximalism in data-driven design is likened with mixed media in art, where more than one medium is used. De facto, the heterogeneity of the data, its sources, and the people who contribute to its creation and curation will insert juxtapositions and collisions. This may not always be desired, and several catastrophic, inconsequential or seemingly random associations should be redacted. However, the ``grain'' of data-driven design \cite{khaled2013pcg} is built on the collision and absurdity of different elements that find their way into the game.

It should be noted that maximalism in the artistic sphere refers to materials or identities of elements within an image, song, or novel. We refer to maximalist game design in that sense, focusing on how game elements originating from different data sources (or transformed in different ways) are visualized, combined and made to interact together, thus not directly opposed to minimalist game design.
\citeauthor{nealen2011towards}'s (\citeyear{nealen2011towards}) minimalist game design encourages removing the unnecessary parts of the design, highlighting the important bits. \citeauthor{sicart2015loops}'s approach (\citeyear{sicart2015loops}) refers to the game loop; minimalist games have a simple core game loop 
which is largely unchanged throughout the game. Sicart uses \emph{Minecraft} (Mojang, 2011) as an example where the simple core loop gather$\rightarrow$craft$\rightarrow$build that remains relevant and unchanged (except from the specific materials worked) throughout the game.

A data-driven game, maximalist in the artistic sense, can also be minimal in the gameplay loop sense. Data Adventures \cite{barros2015dataadventures} has a simple core gameplay loop of traveling to a new location, talking to a non-player character in that location, learning the clue for the next location. Games that we would define as maximalist on the design sense, on the other hand, have the broadest mechanics of options for solving a problem --- e.g. killing a dragon with stealth, magic, followers, swords, fists, poison, etc. in \emph{Skyrim} (Bethesda, 2011) --- or subsystems that are so elaborate or numerous that the player becomes unable to distinguish a core game loop --- e.g. the diverse driving, shooting, spraying, running, etc. minigames in \emph{Saints Row IV} (Deep Silver, 2013) which are the main ways to progress in the game. While certainly data-driven design can offer the latter form of maximalism, e.g. with individual minigames where different forms or sources or data are presented and interacted with in each, not all data-driven games need to have maximalist game loops.

\section{Case Study: Data Adventure Games}\label{sec:datagames}

\begin{figure}[tb]
\subfloat[Data Adventures map screen]{
\includegraphics[width=0.48\columnwidth]{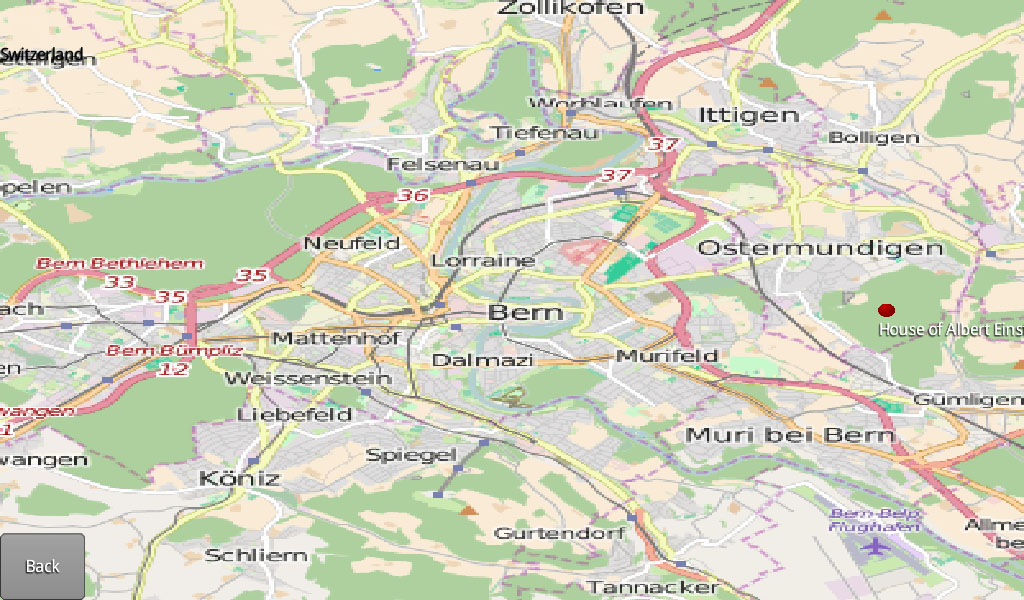}
\label{fig:da1}
}
\subfloat[Data Adventures NPC screen]{
\includegraphics[width=0.48\columnwidth]{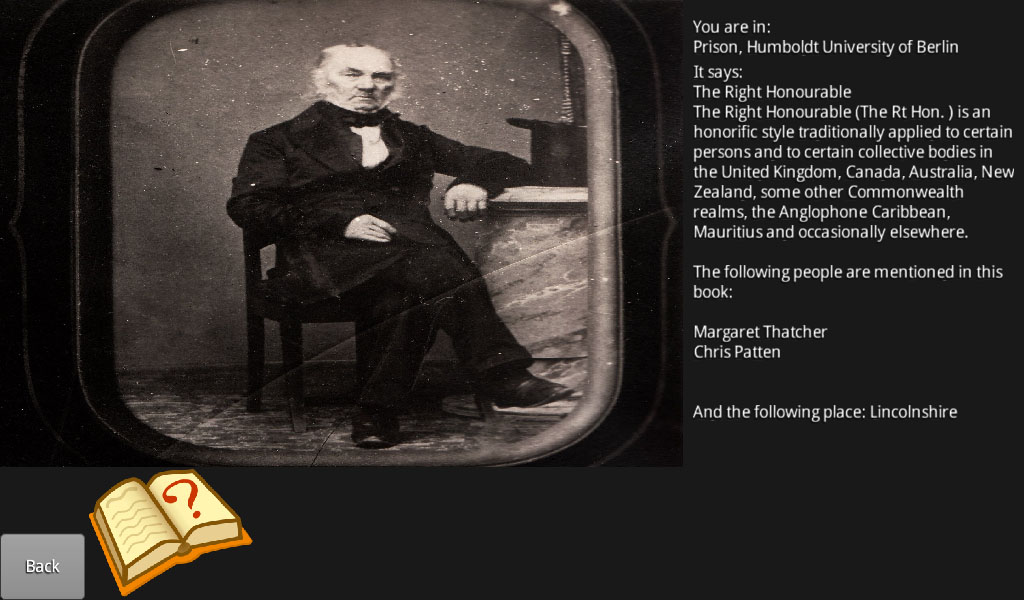}
\label{fig:da2}
}
\\
\subfloat[WikiMystery location screen]{
\includegraphics[width=0.48\columnwidth]{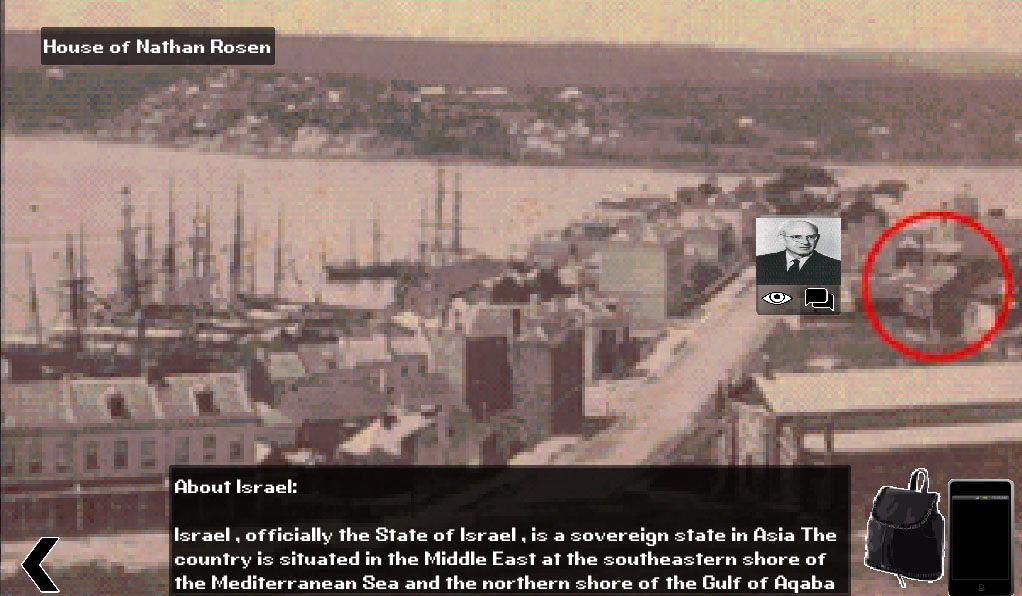}
\label{fig:wm1}
}
\subfloat[WikiMystery accusation screen]{
\includegraphics[width=0.48\columnwidth]{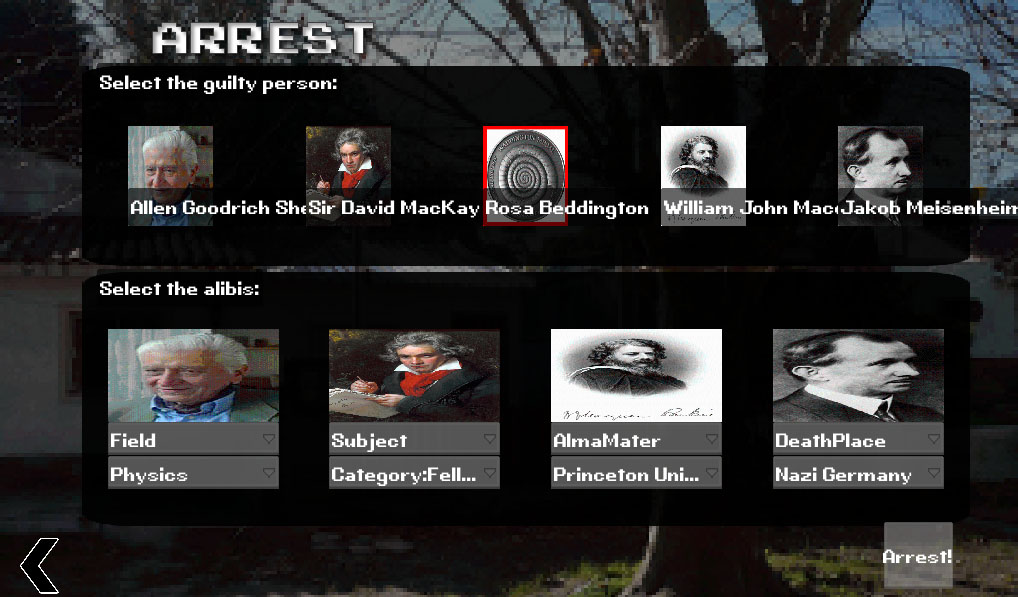}
\label{fig:wm2}
}
\\
\subfloat[DATA Agent dialog screen]{
\includegraphics[width=0.48\columnwidth]{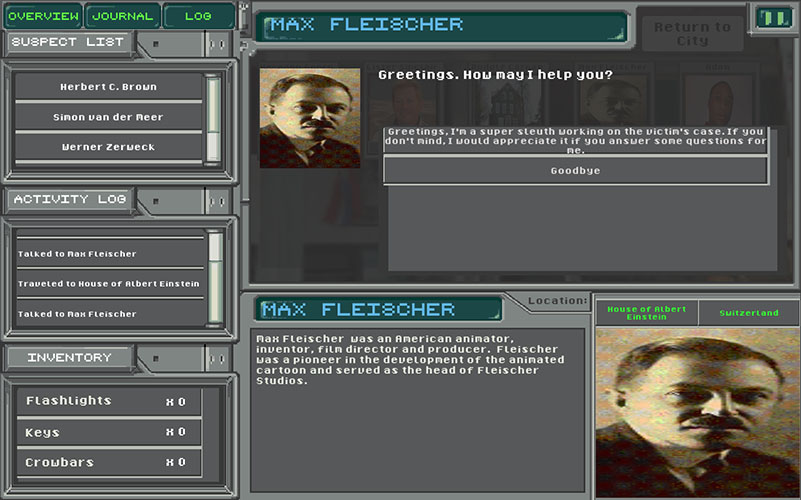}
\label{fig:data1}
}
\subfloat[DATA Agent location screen]{
\includegraphics[width=0.48\columnwidth]{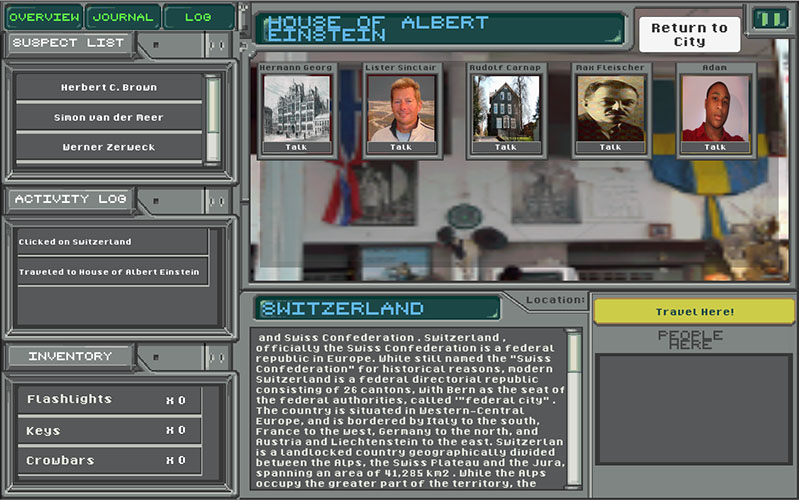}
\label{fig:data2}
}
\caption{Screenshots from the different games in the Data Adventures series. Sources: \protect\cite{barros2016mystery,barros2016playingdata}.}
\label{fig:screenshots}
\end{figure}

The Data Adventures series of game generators exemplify the use of a high volume of data to procedurally generate content \cite{barros2015dataadventures}. The generated adventure games use information gathered from Wikipedia, DBpedia, Wikimedia Commons and OpenStreetMap (OSM) to automatically create an adventure, complete with plot, characters, items and in-game locations. The series consists of three games: Data Adventures \cite{barros2015dataadventures,barros2016playingdata}, WikiMystery \cite{barros2016mystery,barros2018einstein} and DATA Agent \cite{barros2018dataagent}. Each evolved from the previous one, with DATA Agent being the most recent, complex and powerful. Most of the gameplay, however, is the same: a point-and-click interface inspired by ``Where In The World is Carmen Sandiego?'' (Br{\o}derbund Software 1985).

The series' first installment is Data Adventures, an exploration game created from the connections between two Wikipedia articles about specific people. Two Non-Playable Characters (NPCs) are generated representing each of these people. The player receives a quest from the first NPC, asking them to find the second one. To do so, the player has to travel through cities, talking to other generated NPCs and reading books. All information is created from a path linking one article (of the starting NPC) to the other article (of the goal NPC). Figures~\ref{fig:da1} and \ref{fig:da2} show a map screen generated using OSM and a location showing a NPC and a book.

The second game, WikiMystery, plays differently from Data Adventures. On one hand, the game has an arguably more interesting plot, where the player is a detective trying to solve a murder. Additionally, it is generated using only one input: the victim's name. The system finds people related to the victim, forming a pool of possible suspects, and evolves a small list of suspects that are somehow related to each other. It also provides evidence of innocence to any suspect that is, as the name implies, innocent. The player's goal is to find the one suspect which has no evidence of innocence, and arrest him or her. It thus requires that all four pieces of evidence (one per innocent NPC) are collected before the game can be completed. Figures~\ref{fig:wm1} and \ref{fig:wm2} show a location screen and the accusation screen, where the player identifies the culprit and provides evidences of innocence.

In DATA Agent, the player acts as a time-traveler in charge of finding a murder suspect, who went back in time and killed an important person. The game provides a list of suspects, and the player must travel through locations and uncover clues by talking to NPCs or interacting with items, in order to identify which among the suspects is the culprit.
Similar to WkiMystery, DATA Agent's generator is capable of creating a full adventure when given a real person's name. This person becomes the center of the story, as the victim of a murder. Using artificial intelligence techniques over Wikipedia and DBpedia content, the system finds articles related to the person's article, and fleshes out links between suspects and the victim. Every in-game NPC, object, location, dialog or image is created from real information. 
Unlike WikiMystery, there is no evidence of innocence. The game finishes when an a suspect NPC is interrogated by the player and answers wrongly on personal information; the player must have collected the real information during gameplay. NPCs in the game have a much more involved dialog system, and can give information about suspects or about themselves, such as their birth day and occupation, or the reason they were chosen as suspects by the system. Figures~\ref{fig:data1} and \ref{fig:data2} show a dialog screen and an in-game location.


\section{Designing Games for Maximalism}
A major challenge of maximalist game design is deciding what to prioritize. One can shape data in order to fit the game, or modify the game design to better showcase the original data. One can also have data ingrained in the game mechanics, directly affecting gameplay, or show the data in a decorative manner. Maintaining a balance between data transformation to fit other data and the game itself, or staying faithful to the original data while providing an engaging experience is challenging. This section describes two dimensions of maximalist game design: \textit{Data Transformation versus Data Fidelity} and \textit{Functionality versus Decoration}. 

\subsection{Data Transformation versus Data Fidelity}

The tension between data fidelity and data transformation is rooted in the priorities of a maximalist designer: the original game design or the original data. When using open data, designers may wish to adapt that data to the game, or to keep the data as it is and mold the game around it. Extensive data transformations may improve the game experience, but are also susceptible to loss of information or inaccuracies.

Transforming data gives designers more freedom and might be preferred if they have an inflexible idea, or if the data itself is malleable. DATA Agent is such an example of data transformation. In the game, the engine transforms individual facts about separate people into a murder mystery. The facts are also transformed into dialog lines, used by NPCs when prompted by the player. Some facts are altered purposely, in order to ``lie'': the culprit's dialog differs from reality in order to point to the time-agent (and thus, the player) that he or she is guilty. WikiMystery, on the other hand, uses proof of innocence in a similar manner but never misrepresents the actual data: all proofs given by NPCs are true, and the player must memorize them in order to use them in the game's accusation sequence.

On the other hand, designers may instead wish to stay faithful to the original data, molding the game to the data instead. This way, information present in the data is more likely to be clearly presented within game content. 
The rigidity to data restricts what kind of game elements can be used, or forces designers to be creative in their implementations. While less time might be spent cleaning and translating data, more time will likely be spent on raw game and mechanic design. An example can be found in Data Adventures, where data instantiated in the game is sourced from OSM and Wikipedia articles about people, places, and concepts. The designers built a game that could involve all four of those elements: a game where the player travels around the world searching for links to the goal NPC. It introduces some alterations from the original material, but most of it remains unchanged in the game. However, the game lacks a convincing narrative and theme, such as a murder mystery in later installments of the data adventures series. Another example is WikiRace\footnote{http://2pages.net/wikirace.php}, where the game uses Wikipedia to navigate the game.

\subsection{Functional versus Decorative}

Another dimension pertaining to maximalist game design is functionality versus decoration. We define data being \emph{functional} when it has a strong impact on gameplay. If the player does not have to interact with the data, or the data does not impact gameplay in a significant way, then it is \emph{decorative}.

In order to be functional, data can be incorporated in a variety of ways. In DATA Agent, dialog and character names heavily rely on open data. To progress in the game, you must interact with these characters and talk to them. The data is functional, as remembering which NPC (by name) has a certain fact is necessary to identify the culprit. In OpenTrumps \cite{cardona2014open}, the raw mechanics of the game come from open data, as the cards themselves are created from it. The maps in \emph{FreeCiv} generated by \citeauthor{barros2015balanced} (\citeyear{barros2015balanced}) are based on real-world terrain data, and impact gameplay as terrain affects players' city production. 

On the other hand, any data that does not serve a functional purpose is decorative. Data can serve a decorative purpose in many ways. A Rogue Dream \cite{cook2014rogue} uses open data to name player abilities, however the in-game effects of the abilities are not affected by their names or the underlying data. In DATA Agent, city maps and NPC profile images are used as visual stimuli and play no mechanical role in game. \emph{World of Warcraft} (Blizzard, 2004) uses real-world time to create an aesthetic day-night cycle in-game, which has no affect on actual gameplay.

\subsection{Instances of Data-driven Design}

\begin{figure}[tb]
\includegraphics[width=0.95\columnwidth]{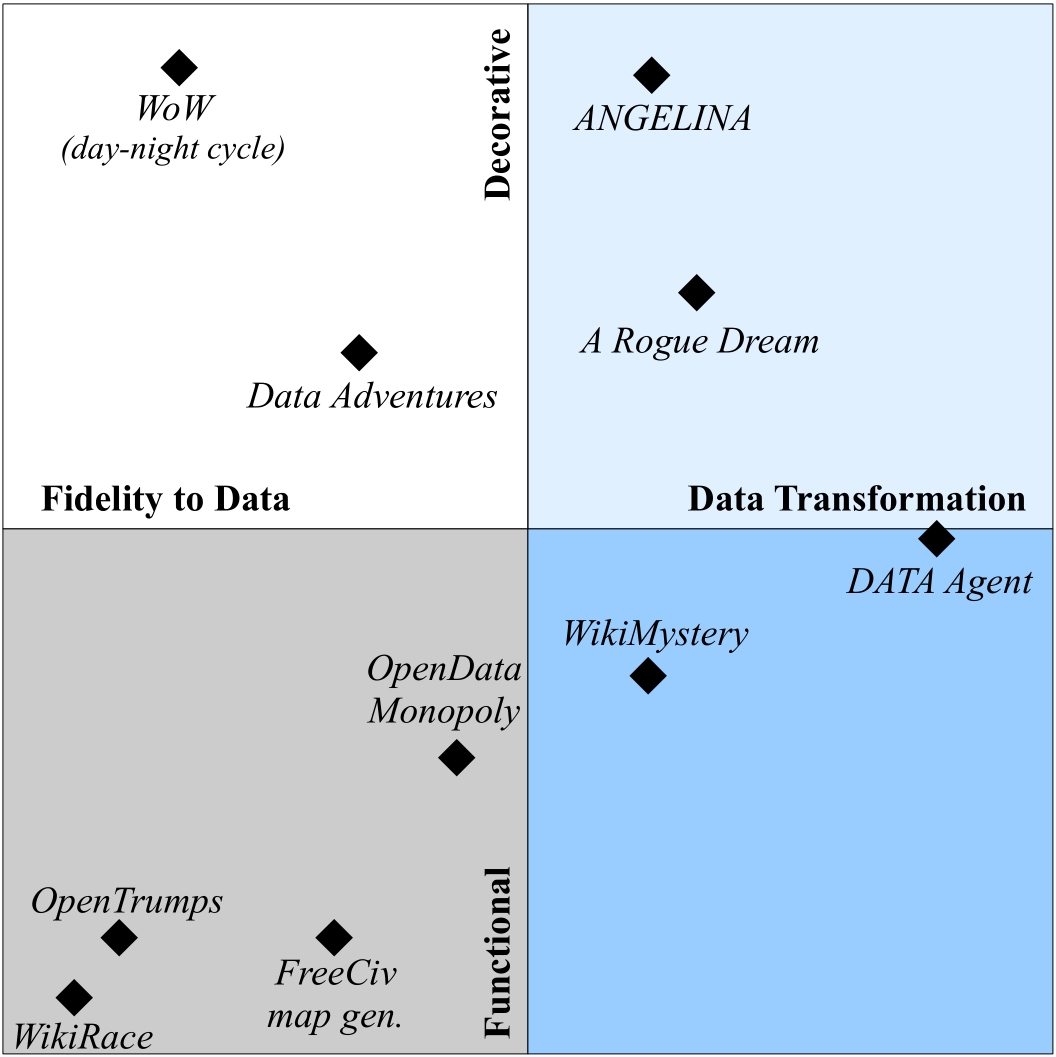}
\caption{Examples of games within the two dimensions.}
\label{fig:maximalism}
\end{figure}

Figure~\ref{fig:maximalism} shows the two dimensions described above. The X-axis represents  Data Transformation versus Data Fidelity, while the Y-axis represents Functional versus Decorative. Games where the goal is to preserve the original data, adapting the game to do so, are on the leftmost side of the figure: WikiRace and OpenTrumps exemplify this. Data in these games is also extremely functional: all mechanics in these games rely on a direct interaction with and understanding of the data (in WikiRace through reading the articles, and OpenTrumps through the values that affect deck superiority). Less faithful to the data, but similarly functional, is the FreeCiv map generation where geographic data is used as-is; resource placement is based on the original data but also adapted (i.e.~transformed) for playability.

Moving upwards along the Y-axis, we find Open Data Monopoly \cite{friberger2012monopoly} and WikiMystery. Both use data in a decorative manner, the former as names for lots on the board and the latter as images, but some of the original data is also functionally translated (e.g. lots prices in Open Data Monopoly and proof of evidence in WikiMystery). On the far end of the Y-axis we have ANGELINA \cite{cook2012aesthetic}, which uses visuals as framing devices but without affecting the core platformer gameplay. The visuals are based on text of a newspaper article: the source data is transformed via natural language processing, tone extraction, and image queries.

\section{Purposes of Maximalist Games}

While we attempt to highlight the principles and directions for designing maximalist game experiences, it is important to consider the purpose of such a game. Maximalist game design is desirable for many different reasons, highlighted in this section. Depending on the purpose, moreover, the design priorities could shift in the spectrum of data fidelity or decoration versus function.

\subsection{Learning}

Modern-day practice sees students of all levels refer to Wikipedia for definitions, historical information, and tutorials. When browsing such knowledge repositories, it is common to access linked articles that may not have been part of the topic of inquiry. Maximalist game design that exploits open sources of knowledge, such as Wikipedia, can be used as a tool for learning in a playful context. One strength found in games is their ability to motivate a player for long periods of time. They also allow several ways to engage players, which can vary based on decisions and learning goals of game designers. Furthermore, games present failure as a necessary event for learning \cite{plass2015foundations}, causing players to explore and experiment more, since failing in game is less consequential than in the real world \cite{hoffman2010motivational}. Studies have also shown that when games immerse the player in a digital environment, they enhance the player's education of the content within the game \cite{dede2009immersive}. Games, unlike raw open data, are adaptive to players' skill level and are the most fun and engaging when they operate on the edge a player's \emph{zone of proximal development} \cite{vygotsky1978interaction}. Thus, we believe players can learn facts within open data during gameplay.

Data-driven maximalist games intended for learning 
can highlight and allow the player to interact with the data playfully. DATA Agent, to a degree, builds on this concept by creating NPCs out of articles about people, whether they are historical or fictional. These NPCs can then answer questions about their birth date or life's work. More relevant to the game progression, each NPC leads to an object (NPC, item or location) about another article, which can be interacted with and is associated to the current NPC somehow. The Data Adventures series was not designed with the explicit purpose of learning in mind; possibly, alternative design priorities could build on more ``educational'' principles. It would be possible, for example, to check for understanding by asking the player questions relating to the data in a diegetic manner --- not unlike DATA agent, where the player must interrogate suspects and cross-check with their own obtained knowledge to detect falsehoods.

Information used to instantiate data should be fairly transparent to the player-learner. Therefore, transformation of data should not be convoluted, and perhaps even textual elements from original articles can be used as ``flavor text''. The veneer between encyclopedic content and game content does not need to be thick, in order to ensure that the right information is provided. In terms of function versus decoration, maximalist games for learning tend to edge closer towards data that influences the outcome of a game session in order to motivate learners to understand and remember the data. Such checks for understanding, however gamified they may be, will have an impact on the success or failure of the game.

\subsection{Data exploration}

Data transformed into interactive game content, forming a consistent whole that goes far beyond the sum of its parts, can allow human users to explore the data in a more engaging way. Data visualization has been used extensively with a broad variety of purposes --- far beyond the ones listed here --- to take advantage of how most humans can more easily think through diagrams \cite{vile1998thinking}. In that vein, gameplay content originating from data can act as a form of highly interactive data visualization. The fact that data from different sources is combined together based on associations imagined by an automated game designer allows players to reflect on the data and make new discoveries or associations of their own. Due to the potential of emotional engagement that games have beyond mere 2D bar plots, the potential for lateral thinking either through visual, semantic, gameplay or emotional associations \cite{scaltsas2013creating} is extraordinary. In order for a game to offer an understanding of the data that is used to instantiate it and allow for that data to be re-imagined, the transformation into game content should be minimal. Examples of data games which already perform such a highly interactive data visualization are BarChartBall \cite{togelius2013barchartball} or OpenTrumps \cite{cardona2014open}. However, a more maximalist approach could benefit games like the above by providing a more consistent storyline and progression, as well as a stronger emotional investment in the data.

\subsection{Contemporaneity}

Automated game design has been always motivated, to a degree, by the desire to create perpetually fresh content. With data-driven design, this can be taken a step further by generating a new game every day. Such a game could be contextually relevant based on the day itself, e.g. building around historical events which happened on this day (such an extensive list can be found on \url{onthisday.com} and Wikipedia) or people who had important personal events on that day (e.g. date of birth, date of death, graduation day). Moreover, the social context can be mined and used to drive the automated design process by including for instance trending topics of Twitter or headlines of today's newspapers. Early examples of such data-driven process have been explored for example by ANGELINA \cite{cook2012aesthetic} which used titles and articles from The Guardian website and connected them with relevant visuals and the appropriate mood. It is expected that a more maximalist data-driven design process would strengthen the feeling of contemporaneity by including more data sources (i.e.~more data to transform) or stronger gameplay implications (i.e.~broader transformations and functional impact).

Contemporaneity can make games generated on a specific day appealing to people who wish to get a ``feel'' for current issues but not necessarily dig deeply. On the other hand, the plethora of games (30 games per month alone) and the fact that each game is relevant to that day only could make each game itself less relevant. Contemporaneity and the fleeting nature of daily events could be emphasized if each game was playable only during the day that it was produced, deleting all its files when the next game is generated. 
This would enhance the perceived value of each game, similarly to \emph{permadeath} in rogue-like games as it enhances nostalgia and the feeling of loss when a favorite gameworld is forever lost.

Any maximalist game could satisfy a contemporaneity goal, but such games can be more amenable to data transformation. For example, data could be transformed to more closely fit the theme of the day, e.g.~query only female NPCs on International Women's Day.
Contemporaneous data can be functional (to more strongly raise awareness of issues) but can also easily be decorative, e.g. giving a snowy appearance to locations during the Christmas holidays.

\subsection{Personalization}

When game content is generated from data, it is possible to highlight certain bits of information. When the game takes player input as part of the data selection process, it personalizes their experience. If player information is available in the form of interests, important personal dates such as birthdays, or even social networks, the potential data sources that can be selected to form the game can be narrowed down. Presenting game content which is personally relevant (e.g. adventures with NPCs based on people living before Christ for an archeology student), or contextually relevant (such as solving the murder of an NPC born on the player's birthday) could contribute to a more engaging experience. It might also be possible to tailor the game's source repositories based on such personal interests. There are numerous online wikis, most of which follow a common format; therefore a user can implicitly (via personal interests) or explicitly (by providing a direct URL) switch search queries of a data-driven maximalist game to a specific wiki of choice.
 
\subsection{Opinion \& Critique}

Often designers want to make a statement through their games. For instance, Game-o-matic \cite{treanor2012gameomatic} creates games from manually defined associations (as \emph{micro-rhetorics}). \emph{September 12th: A Toy World} (Newsgaming 2003) makes a political statement about the futility of America's War on Terror. Open data could similarly be used in a game to critique some aspect of culture by adding a weight of relevance and realism. For instance, a game such as \emph{September 12th} could use the real map or skyline of Baghdad, or data on daily deaths in Iraq, to instantiate the challenge of the game. Similarly, if designers wish to critique the unprofessional use of social media in the White House, one could use real tweets to form dialog lines rather than generating them as in DATA Agent \cite{barros2018dataagent}. 

\subsection{Entertainment}

Ostensibly, all games have entertainment as a (primary or secondary) purpose. This includes maximalist games, even if they have an additional purpose as listed in this paper. It is meaningful therefore to investigate what data-driven maximalist design has to offer to the entertainment dimension of any such game. Since maximalism ---as we define it--- does not necessarily apply to the mechanics of a game, a more relevant dimension is the end-user aesthetic that such games facilitate, following the mechanics-dynamics-aesthetics framework of \citeauthor{hunicke2004mda} (\citeyear{hunicke2004mda}). Data-driven maximalist games primarily enhance the aesthetic of \emph{discovery}, similarly to data exploration via such a game, and \emph{expression} if it can be personalized to a user based on provided input such as birthday, hometown or interests. In many ways, data-driven games can enhance the aesthetic of \emph{fantasy} by using and transforming real-world information. DATA agent, for example, describes an alternate history setting where a famous historical figure has been murdered (often by colleagues). The fantasy aesthetic is further enhanced by having a player take the role of a detective traveling through time and space to interrogate suspects. Other possible aesthetics that can be enhanced through data are \emph{sensation} if the data comes from sources of high quality video, audio, or visuals (e.g. paintings of the National Gallery of London), or \emph{fellowship} if the data comes from other users (e.g. anonymous users' trending tweets or social media postings of the player's friends). Evidently, games geared primarily towards entertainment can be fairly flexible in terms of data transformation, and can adapt the data to the intended game mechanics and game flow. While data can act as a decoration in such games (if intended to enhance the sensation aesthetic), in general games intended primarily for entertainment are fairly focused in the mechanics and feedback loops, and thus data would primarily be transformed into functional elements.

\subsection{Human Computation}

Presenting hand-picked results from a vast database in an engaging, playful way is not only relevant for humans to consume. The human-computer interaction loop can be closed if human users provide feedback on the quality of the data itself. This human feedback can be used internally by the game, adapting its criteria in order to avoid unwanted data repositories, queries, associations or transformations made to the data. For instance, a future DATA agent version could re-compute the set of suspects for the next games (removing one or more suspects from the pool of possible suspects) if a player provides negative feedback explicitly (e.g. via a `report' button) or implicitly (e.g. by being unable to solve the mystery). More ambitiously, the positive or negative feedback of players engaging with the playable ---transformed--- data can be fed back to the source repositories which instantiated the game. This can allow for instance misinformation found in Wikipedia to be flagged, alerting moderators that either a human error (e.g. a wrong date or a misquote) or malformed data (e.g. unreadable titles) exists and must be corrected. Whether these corrections should be made by an expert human curator, or directly based on player interactions with the game could be a direction for future research.

\section{Issues with Data-Driven Game Design}

Accomplishing \emph{good} data-driven maximalist game design is a challenge. While the previous sections presented ways of doing so, there are still many implementation- or game-specific details which affect the design process. Beyond the core challenge of a good game design, there are several peripheral challenges to the design task itself which however spring from the practice of data-driven design. We elaborate on those peripheral challenges here.

\subsection{Legal \& Ethical Issues}
Any software which relies on external data that it cannot control may be prone to legal or ethical violations. Privacy of personal information may be a concern for a game generated from the social media profile of a user, especially if that game can then be played by a broader set of people. Using results from Google Images may lead to direct infringements of copyrights; using results from models built from text mining, on the other hand, may or may not result in such copyright infringements depending on whether the model returns actual copyrighted material. The issue of copyright becomes more complex when the data is transformed: relevant to data mining, a judge has ruled for fair use for Google Books as ``Google Books is also transformative in the sense that it has transformed book text into data for purposes of substantive research, including data mining and text mining in new areas'' \cite{sookman2013google}. One can only assume that transformations of data into game content, depending on the fidelity to the original data and the purpose (e.g. data exploration and education), would make for a clearer case of fair use.

Game content built on fair use or open data combined into an interactive experience may lead to unexpected issues. This is especially true in cases where the player has sufficient agency to interpret or act upon content of high fidelity with the original data in an open-ended fashion: consider, for example, a violent shooter game where opponents' visual depictions (3D models or faces) are those of Hollywood celebrities. Even in Data Adventures, where player interaction is fairly ``curated'', a generated game featured solving the murder of Justin Bieber \cite{barros2016mystery}. Apart from the fictional narrative of a popular celebrity's death, the game identifies another celebrity as the murderer: both of these decisions may cause concern to highly visible people (be they depicted murdered, murderers, or suspects). A disclaimer that the game characters are fictional can only alleviate that much of the ethical responsibility of game designers for such data-driven games.

\subsection{Misinformation \& Bias}

Connected to the concerns of misrepresenting contemporary or historical celebrities are the inherent issues of error in the source data. Before data is transformed into game content, open repositories that can be edited by anyone can be saturated by personal opinion and perhaps deliberate misinformation. As noted previously, not all data provided by different stakeholders in the information age are factual; this may be more pronounced in certain repositories than others. Beyond deliberate misinformation, an inherent bias is also present even in ``objective'' data. For example, algorithms for Google query results or image results are based on machine learned models that may favor stereotypes (based on what most people think of a topic). 
Even though WikiMystery uses what we arguably consider ``objective'' repositories such as Wikipedia, the 8 most popular locations in 100 generated games were in North America \cite{barros2018einstein}, pointing to a bias of the articles or the DBpedia entries chosen to be digitized. Other cases where misinformation may arise is when different content is combined inaccurately: examples from the Data Adventures series include cases where an image search for a character named Margaret Thatcher resulted in an image of Aung San Suu Kyi \cite{barros2016playingdata}. When data-driven design uses social network data such as trending topics on Twitter, then the potential for sensitive or provocative topics to be paired with inappropriate content or combined in an insensitive way becomes a real possibility. If data-driven maximalist games are intended towards critique or opinion, the misinformation or misappropriation could be deliberately inserted by a designer (by pairing different repositories) or accidentally introduce a message that runs contrary to the intended one.

\section{Outlook}

Maximalist game design encourages creation through reuse and combination. If one imagines its most powerful form, it would likely involve taking any mixture of information, pouring it into any game content cast, and reveling in its results. It would provide a freedom to interact with any data in the best, most personalized way possible. 

Current PCG techniques allow for unlimited playability for a large variety of games. However, they can lack a level of contemporaneity and relevance that could be provided by open data. Additionally, research has suggested that concepts can be effectively learned through gameplay \cite{dede2009immersive}. Using games as a method of interacting with open data may create a novel way for learning about the data in a fun way. Rather than use Wikipedia to learn about specific people and places for the first time, players could play games where they can talk to these people and visit these places.

Open data is available to all, to create as well as consume. Sometimes the data is inaccurate. The idea of visualizing this information in any form can provide means to ``debug'' the original data, in a more engaging way than just browsing Wikipedia or poring through a massive database.



\section{Conclusion}\label{sec:conclusion}

This paper discussed an approach to game design inspired by the notion of maximalism in the arts. It encourages the reuse and combination of heterogeneous data sources in the creative design process. Maximalist game design embraces the generation of game content using different data sources, re-mixing them in order to achieve something new.

We drew from our experience with the Data Adventures series to propose a mapping of the maximalist game design space along two dimensions, \textit{data transformation versus data fidelity} and \textit{functionality versus decoration}. The former focuses on the extent that the data is transformed from its original form, while the latter refers to the actual role of the data in the game. Additionally, we described how maximalist game design can serve different purposes in the design process and which tradeoffs emerge from each purpose. Finally, we highlight issues and ethical concerns that may arise from and in maximalist games.  

\section*{Acknowledgements}

Gabriella Barros acknowledges financial support from CAPES and Science Without Borders program, BEX 1372713-3. Antonios Liapis has received funding from the European Union's Horizon 2020 research and innovation programme under grant agreement No 693150.

\bibliographystyle{iccc}
\small
\bibliography{maximalism}

\end{document}